\documentclass[aps,prb,twocolumn,superscriptaddress,showpacs,preprintnumbers,amssymb]{revtex4-1}
\usepackage{graphicx }
\usepackage{dcolumn}
\usepackage{bm}
\usepackage{color}
\newcommand{\YNP}{YbNi$_{4}$P$_{2}$}
\newcommand{\YNPA}{YbNi$_{4}$(P$_{1-x}$As$_{x}$)$_2$}
\newcommand{\mub}{$\mu_{\mathrm{B}}$}
\newcommand{\TC}{$T_{\mathrm{C}}$}
\newcommand{\xc}{$x_{\mathrm{c}}$}
\begin{document}

\preprint{APS/123-QED}

\title{Magnetic order and spin dynamics across a ferromagnetic quantum critical point: $\mu$SR investigations of \YNPA}
\author{R. Sarkar}
\altaffiliation[]{rajibsarkarsinp@gmail.com}
\author{J. Spehling}
\author{P. Materne}
\affiliation{Institute for Solid State Physics, TU Dresden, D-01069 Dresden, Germany}
\author{H. Luetkens}
\author{C. Baines}
\affiliation{Laboratory for Muon-Spin Spectroscopy, Paul Scherrer Institute, CH-5232 Villigen, Switzerland}
\author{M. Brando}
\affiliation{Max-Planck Institute for Chemical Physics of Solids, 01187 Dresden, Germany}
\author{C. Krellner}
\affiliation{Institute of Physics, Goethe University Frankfurt, Max-von-Laue-Strasse 1, 60438 Frankfurt am Main, Germany}
\author{H.-H. Klauss}
\affiliation{Institute for Solid State Physics, TU Dresden, D-01069 Dresden, Germany}
\date{\today}
\begin{abstract}
In the quasi-1D heavy-fermion system \YNPA\ the presence of a ferromagnetic (FM) quantum critical point (QCP) at \xc\ $\approx 0.1$ with unconventional quantum critical exponents in the thermodynamic properties has been recently reported. Here, we present muon-spin relaxation ($\mu$SR) experiments on polycrystals of this series to study the magnetic order and the low energy 4$f$-electronic spin dynamics across the FM QCP. The zero field $\mu$SR measurements on pure \YNP\ proved static long range magnetic order and suggested a strongly reduced ordered Yb moment of about 0.04\,\mub. With increasing As substitution the ordered moment is reduced by half at $x = 0.04$ and to less than 0.005\,\mub\ at $x=0.08$. The dynamic behavior in the $\mu$SR response show that magnetism remains homogeneous upon As substitution, without evidence for disorder effect. In the paramagnetic state across the FM QCP the dynamic muon-spin relaxation rate follows 1/$T_{1}T\propto T^{-n}$ with $1.01 \pm 0.04 \leq n \leq 1.13 \pm 0.06$. The critical fluctuations are very slow and are even becoming slower when approaching the QCP.
\end{abstract}
\maketitle
Quantum criticality is a central topic in the field of strongly-correlated electron physics.~\cite{{Focus-QPT-2008-1},{Focus-QPT-2008-2},{Focus-QPT-2008-3},{Focus-QPT-2008-4}} In $d$- and $f$-electron based metallic systems close to an antiferromagnetic (AFM) ground state, quantum critical points (QCPs) are well established from both, the experimental and theoretical point of view. On the other hand, the existence of a ferromagnetic (FM) QCP in metals is controversially discussed.~\cite{Brando-2015} In fact, in clean 2D and 3D itinerant ferromagnets theoretical studies deny the presence of a FM QCP and the second order FM quantum phase transition (QPT) becomes first order before reaching the QCP.~\cite{Belitz-1999,Chubukov-2004} This has been observed in several $3d$-electron ferromagnets like MnSi~\cite{Pfleiderer-1997}, ZrZn$_{2}$~\cite{Uhlarz-2004} or in U-based systems~\cite{Aoki-2011}. Other theories and experimental studies demonstrate that an inhomogeneous magnetic phase forms in between the paramagnetic and the homogeneous FM phase.~\cite{Belitz-1997,Conduit-2009,Karahasanovic-2012,Brando-2008,Kotegawa-2013} In other systems the formation of disordered phases, e.g. quantum Griffiths phase or spin-glass-like state, have been observed instead of a FM QCP.~\cite{Westerkamp-2009,Ubaid-Kassis-2010,Lausberg-CeFePO-2012} In $5f$-electron based metals like, UGe$_{2}$~\cite{Taufour-2010} or UCoGe~\cite{Huy-2007}, it was observed that FM QCPs have been avoided by the formation of superconducting phases. In summary, evidence for the existence of a FM QCP in metallic systems has been lacking so far. However, the recent observation of power-law divergence of the Gr\"uneisen ratio in the Kondo-lattice ferromagnet \YNPA\ at the critical concentration \xc\ $\approx 0.1$~\cite{Steppke-2013}, which proves the existence of a QCP,~\cite{Zhu-2003} has reopened the discussion. This material has quasi-1D crystalline and electronic structures,\cite{Krellner-2011} and a strong Kondo interaction. 

1\,D fluctuations are expected to supress the mechanism which induces the first order transition. Thus, it seems that in such materials FM QCPs could be observable. It became therefore of relevant importance to study the spin dynamics in \YNPA\ across its  FM QCP.

The heavy-fermion metal \YNP\ possesses a very low Curie temperature \TC\ = 0.15\,K and a small ordered Yb moment due to the strong Kondo effect with a Kondo temperature of 8\,K.~\cite{Krellner-2011,Spehling-2012,Steppke-2013} The quasi-1D crystal structure is due to Yb and Ni-tetrahedra chains along the $c$-axis of the tetragonal unit cell.~\cite{PIVAN1989285} Susceptibility measurements between 50 and 400\,K indicate clear Curie -Weiss behaviour with an effective moment close to the value of trivalent Yb.~\cite{Krellner-jpcm-2012} A recent resonant X-ray emission spectroscopy (RXES) study reveal a low-temperature valence of Yb$^{+2.988}$.~\cite{Kummer-2016} 
 Non-correlated band-structure calculations found a quasi 1D electronic structure and a non-magnetic Ni$_4$P$_2$ sublattice.~\cite{Krellner-2011} The absence of Ni-related magnetism in \YNP\ is further supported by the fact that LuNi$_4$P$_2$ presents a non-magentic ground state.~\cite{Kliemt-2016} In \YNP\, the presence of strong FM correlations was seen in NMR experiments above 2\,K,~\cite{Sarkar-2012,Baenitz-2013} and in inelastic neutron scattering.\cite{Huesges-JPCM-2015} $\mu$SR experiments were already performed on pure \YNP\ and have proved homogeneous static magnetic order with an ordered moment of about 0.04\,\mub. Above \TC\ the muon spin polarization obeys a time-field scaling relation indicating cooperative critical spin dynamics, with the dynamic spin-spin autocorrelation function following a power law behavior. These results suggest that the NFL behavior above \TC\ is induced by quasi homogeneous critical spin fluctuations.~\cite{Spehling-2012}

\begin{figure}[t]
\includegraphics[width=\columnwidth]{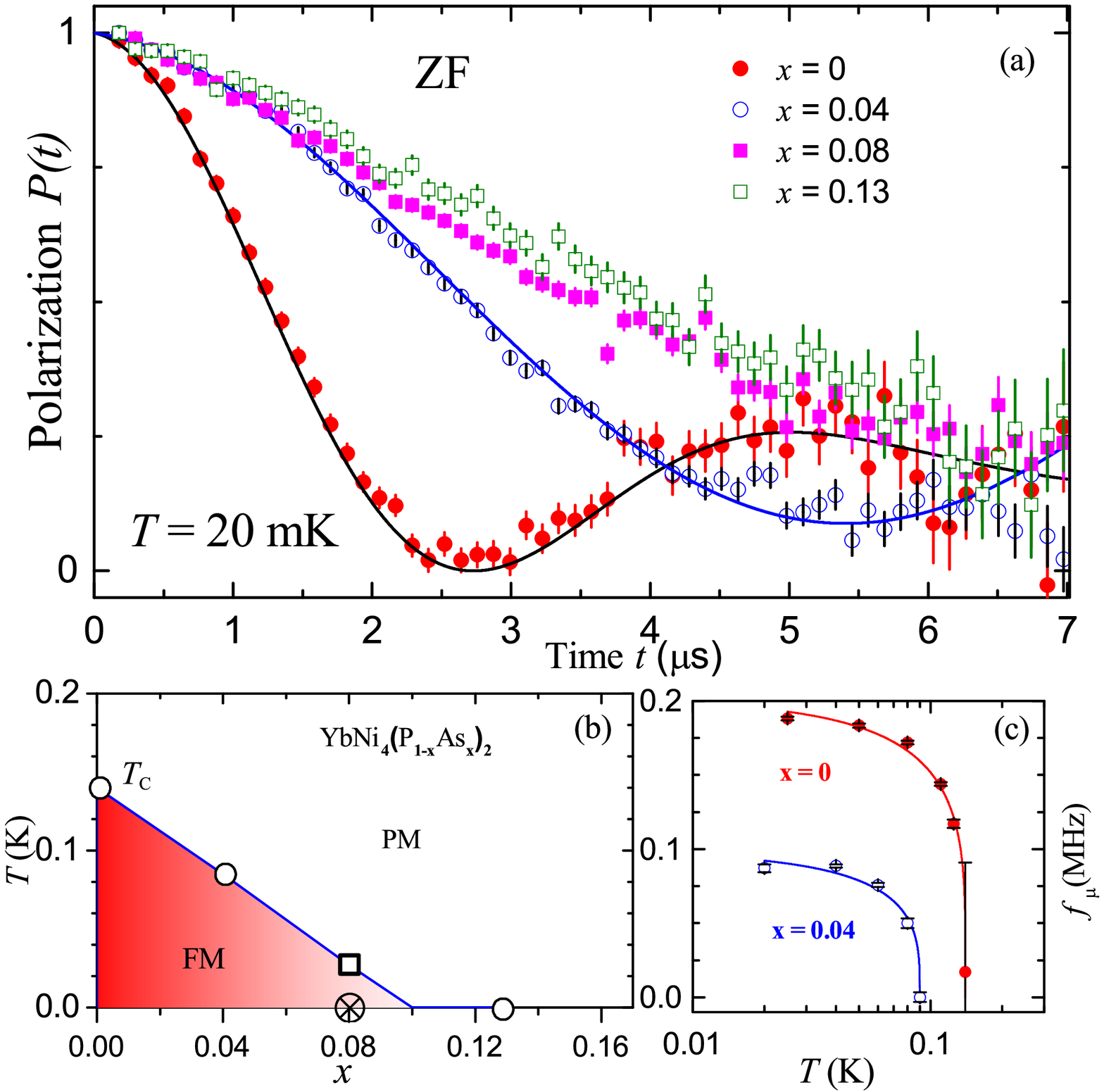}
\caption{\label{fig:fig1}a). Panel (a) shows the time spectra for all As concentrations at $T = 20$\,mK. The solid lines denote the theoretical simulation as explained in the main text. Panel (b) outlines the phase diagram of \YNPA\ as determined in Ref.~\onlinecite{Steppke-2013} ($\Box$) and compared with points derived from our $\mu$SR investigations ($\bigcirc$, $\bigotimes$). Panel (c) depicts the fitted magnetic order parameter ($\mu$SR frequencies). The solid line is a fit to the phenomenological function $f_{\mu}$ (see main text).}
\end{figure}

With a small amount of As substitution for P it is possible to tune \TC\ continuously to zero.~\cite{Steppke-2013}
In fact, for $x = 0.08$, \TC\ is suppressed below 25\,mK and the sample with $x = 0.13$ is non-magnetic (see Fig.~\ref{fig:fig1}b). For $x = 0.08$, both the specific heat $C(T)/T$ and the volume thermal expansion coefficient $\beta(T)/T$ diverge according to power laws with unconventional exponents, namely, $T^{-0.43}$ and $T^{-0.64}$, respectively. Therefore the Gr\"uneisen ratio $\Gamma(T)=\beta(T)/C(T)$ diverges as $T^{-0.22}$ indicating the presence of a nearby FM QCP for $x \approx$ \xc. 

Here, we present muon-spin relaxation ($\mu$SR) studies on polycrystalline \YNPA\ with $x$\,=\,0, 0.04, 0.08, and 0.13 down to $T = 20$\,mK. We show 
that with increasing As substitution the ordered moment decreases continuously to values smaller than 0.005\,\mub\ at $x \approx$ \xc. In the paramagnetic state across the FM QCP the dynamic muon-spin relaxation rate 1/$T_{1}T \propto T^{-n}$ with an exponent value that  decreases across the FM QCP from 1.13\,$\pm$\,0.06 for $x\,=\,0$ to 1.01\,$\pm$\,0.04 for $x = 0.13$. This behavior can not be explained by any existing theory of FM quantum criticality. Finally, from the measured spin autocorrelation time, we find that the spin fluctuations are very slow and becomes even slower for $x \rightarrow$ \xc.

The powder samples were prepared by milling single crystals of the respective concentration. 
The single crystals were grown by a modified Bridgman-method from a Ni-P self-flux in a closed Ta container and are from the same batches as used in Ref. \onlinecite{Steppke-2013}. The excess flux was removed by centrifuging at high temperatures followed by etching in diluted nitric acid, details are described in Ref. [\onlinecite{Krellner-2016}]. The phase purity of the used samples was analyzed by powder x-ray diffraction and magnetization measurements and we found phase pure samples beside small residuals of the flux of order 2\%. 
The positive spin-polarized muons are implanted into the sample in a $\mu$SR experiment, and the time evolution of the muon-spin polarization, $P(t)$, is monitored by detecting the asymmetric spatial distribution of positrons emitted from the muon decay.~\cite{schenck1985muon,yaouanc2011muon} Zero magnetic field (ZF) and longitudinal applied magnetic fields (LF) with respect to initial muon-spin polarization were used to perform $\mu$SR experiments. All experiments were carried out at the $\pi$M3 beam line at the Swiss Muon Source at the Paul-Scherrer-Institute, Switzerland. For thermal contact, the samples were glued on a Ag plate giving rise to a time and temperature independent background signal due to muons that stopped in the Ag plate (about 20-30 \%). The $\mu$SR time spectra were analyzed using the free software package MUSRFIT.~\cite{Suter201269}  

Fig.~\ref{fig:fig1}a represents ZF $\mu$SR time spectra for $x =0 $, 0.04, 0.08 and 0.13 collected at $T = 20$\,mK. The plot clearly indicates that the spontaneous frequency of oscillation associated with the static ordered Yb magnetic moment decreases with increasing As concentration. To extract the magnetic order parameters in the ordered state, the muon asymmetry data were modeled by the following function (solid lines in Fig.~\ref{fig:fig1}) \cite{Spehling-2012,Barsov-1991,Kornilov-1991}
\begin{eqnarray}\label{eq:equation1}
aP(t) &= &a_s(\ \frac{1}{3}+ \frac{2}{3}[\cos(2\pi f_\mu t) \nonumber
\\  & &-\frac{\sigma^{2}t}{2\pi f_\mu}\sin(2\pi f_\mu t)]e^{-\frac{1}{2}\sigma^2t^2})\ 
+ a_{bg}e^{-\lambda_{bg}t}.
\end{eqnarray}
Here, $f_{\mathrm{\mu}}$, $\sigma$ and $\lambda_{bg}$ are the muon-precession frequency, the Gaussian field width of the internal field distribution and the constant background contribution, respectively. The 2/3 oscillating, and the 1/3 non-oscillating terms originate from the spatial averaging in polycrystalline samples, where 2/3~(1/3) of the internal magnetic field components are directed perpendicular (parallel) to the initial muon spin, causing a precession (no precession) of the muon spin. The observation of a 2/3 and 1/3 sample signal fractions, and the Gaussian relaxation of $P(t)$ below $T_{\mathrm{C}}$ proof that 100 \% of the sample volume shows static magnetic ordering of the magnetic moments not only for $x\,=\,0.0$ but also for $x\,=\,0.04$. More detailed description of Eq.~\ref{eq:equation1} can be found for example in Refs.~\onlinecite{Spehling-2012,Barsov-1991,Kornilov-1991}. 

Spehling \textit{et al.} have analyzed a number of potential muon stopping sites, but only considered high symmetry positions.~\cite{Spehling-2012} Looking at the distribution of atoms within this structure type we suspect the muon to sit on a lower symmetry position near the 4\,$c$ site, and therefore calculated the size of the ordered moment from the field expected at this site. From the measured $\mu$SR frequency value $f_\mu = 0.188$\,MHz at $T = 20$\,mK, a local internal field at the muon site of $B_\mathrm{local} = 13.87$\,G can be determined, using the relation $B_{\mathrm{local}} = 2\pi\times$f$_{\mu}/\gamma_\mu$, where $\gamma_\mu = 2\pi\times$13.55 kHz/G is the muon gyromagnetic ratio. On the other hand the fitted $\mu$SR frequency at 20 mK is 0.09 MHz for $x$~=~0.04, which corresponds to an internal field value of 6.642\,G. 
 The values are listed in Table 1. With increasing As substitution the ordered moment is reduced by half in $x = 0.04$, and at $x \approx 0.08$, the ordered moment becomes smaller than 0.005\,\mub.

\begin{figure}[t]
\includegraphics[scale=0.5, trim= 10 10 20 110]{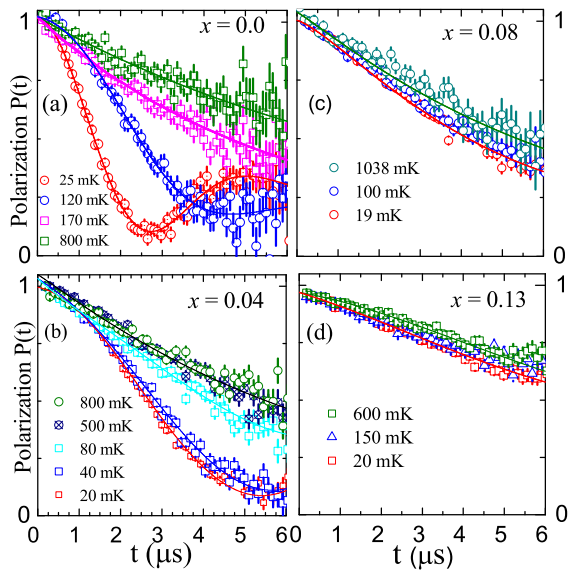}
\caption{\label{fig:fig2} (a)-(d) represent the muon-spin polarization $P(t)$ for $x$ = 0, 0.04, 0.08 and 0.13 at selected temperatures.}
\end{figure}

Figure~\ref{fig:fig2} shows the ZF $\mu$SR time spectra for \YNPA\ with $x = 0$, 0.04, 0.08, and 0.13 concentrations at representative temperatures. A spontaneous $\mu$SR frequency is observed in the time spectra below $T = 140$\,mK, and $T = 90$\,mK for As concentrations $x = 0 $ and 0.04, respectively (Figs.~\ref{fig:fig2}~a,b). This proves the onset of static long-range magnetic ordering of the Yb moments. For $x = 0$ and 0.04, above the magnetic ordering the $\mu$SR time spectra reveal a dominant exponential behavior.
Here, a stretched exponential function $P(t)$~=~$\exp{(-\lambda t)}^{\beta}$ was used to model $P(t)$ indicating that the muon-spin relaxation is due to the dynamics of electronic moments. $\lambda = 1/T_{1}$ is the generalized muon-spin relaxation rate, and $\beta$ is the exponent. The exponent $\beta$ is the measure of the homogeneity of the system. Only slight enhancement (value varies 1-1.15) of $\beta$ is observed towards low temperatures suggesting that the system is homogeneous. The same functional form was used to describe $P(t)$ for $x$~=~0.08. ~\cite{{MacLaughlin-time-field-scaling},{Michor2006640}}
In comparison to $x$~=~0 and 0.04, a clear development of magnetic order parameter is not observed for $x$~=~0.08 down to $T = 19$\,mK (see Fig.~\ref{fig:fig2}c). However, the $\mu$SR time spectra for $x = 0.08$ look very similar to that of $x = 0$ and 0.04 above $T > 140$\,mK and 90\,mK, respectively. Given that for the $x = 0.08$, $T_C$\,=\,28\,mK was observed in $\chi_{AC}$,  
these $\mu$SR observations suggest that the static ordered magnetic moment is too small to be detected in the $\mu$SR time window which therefore sets an upper limit for the ordered Yb moment of $\mu_B$ $\leq$ 0.005 $\mu_\mathrm{B}$.~\cite{Steppke-2013} This is because the frequency of oscillations related to static magnetic order is too small to be captured in the current $\mu$SR time window. Another possibility, the sample pellet did not reach the temperature of the mixing chamber.

In Fig.~\ref{fig:fig1}c, the fitted $\mu$SR frequencies, $f_{\mathrm{\mu}}$, are plotted together with a phenomenological fit of the function $f_{\mathrm{\mu}}(T)=f_{\mathrm{\mu}}(0)(1-\frac{T}{T_\mathrm{C}})^{n^{\prime}}$ (solid lines) for $T < T_\mathrm{C}$, where $n^{\prime}$ is the exponent. 
From our fit we find that $n^{\prime}$ slightly increases from 0.21(2) to 0.28(5) with increasing As concentration.
However the very low data point density for the $T$-range 0.6 $\leq T/T_\mathrm{C} \leq $ 1 prevents a precise estimation of exponents.

For the sample with $x = 0.13$, all $\mu$SR time spectra reveal a pronounced Gaussian functional form instead of a stretched exponential (see Fig.~\ref{fig:fig2}d) as expected for a non-magnetic ground state in which the muon-spin polarization, $P(t)$, is due to the interaction with nuclear dipole moments only. $P(t)$ is best modeled by the following functional form:
\begin{equation}
\label{eq:equation2}
aP(t)=a_s\left[\frac{1}{3}+ \frac{2}{3}(1-\sigma^{2}t^2)e^{-\frac{1}{2}\sigma^2t^2}\right]e^{-\lambda t} + a_{bg}e^{-\lambda_{bg}t},
\end{equation}
where $\sigma$ is the Gaussian field width and $\lambda$ is the muon-spin relaxation rate.    
In a simple paramagnet, $P(t)$ is due to static randomly oriented nuclear dipole moments with a $T$-independent Gaussian relaxation rate $\sigma$, theoretically described by the Kubo-Toyabe function,\cite{PhysRevB.20.850-Hayana} which is the bracketed term in Eq.~\ref{eq:equation2}. An additional exponential term, $\exp(-\lambda t)$, is introduced to account for the extra muon relaxation due to electronic moments, where $\lambda$ is the dynamic muon-spin relaxation rate due to fluctuations of the electronic moments. At all temperatures, the data were analyzed using a fixed $\sigma=0.1316$ MHz determined at high temperatures in the PM state where $P(t)$ is due to nuclear moments only. 

\begin{figure}[t]
\includegraphics[scale=.3]{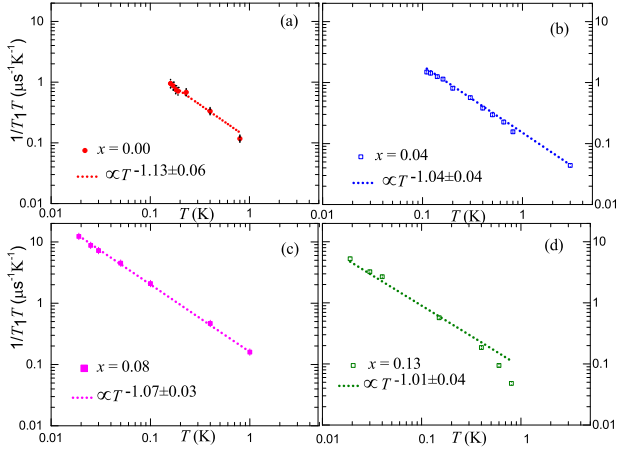}
\caption{\label{fig:fig3}
Temperature dependence of the dynamic muon-spin relaxation rate $\lambda/T = 1/T_{1}T$ for different As concentrations in the paramagnetic state above the ordering temperature. Dotted lines indicate the power-law behavior $1/T_{1}T \propto T^{-n}$. The exponent $n$ slightly decreases from 1.13\,$\pm$\,0.06 to 1.01\,$\pm$\,0.04 with increasing $x$.}
\end{figure}

Figure~\ref{fig:fig3} shows the fitted $\lambda/T$ ($= 1/T_1T$) values as a function of $T$ for $x$~=~0, 0.04, 0.08 and 0.13. An increase of $\lambda/T$ is observed with lowering temperature following a power law behavior $1/T_{1}T \propto T^{-n}$. The exponent value decreases slightly with increasing $x$ from 1.13\,$\pm$\,0.06 for $x = 0$ to 1.01\,$\pm$\, for $x = 0.13$. It should be noted that for the compound $x =0.13$, a different functional form was used to describe the experimental data. The fact that $1/T_{1}T \propto T^{-n}$ with a nearly same exponent for all concentrations is consistent with the observation that the temperature dependencies of the thermodynamic quantities, $C(T)/T$ and $\beta(T)/T$, show the same power-law exponent in the PM region of the phase diagram across the FM QCP. This indicates that disorder does not play a relevant role. For the sample with  $x = 0.08$, located closest to \xc, $1/T_{1}T \propto T^{-1.07\pm0.03}$ indicating that 1/$T_1$ is approximately $T$ independent .

Itinerant ferromagnetism is often discussed on the basis of spin fluctuation theories.~\cite{Moriya-1985} Within the self-consistent renormalization (SCR) theory, the spin-lattice relaxation rate $1/T_{1}T$ in itinerant ferromagnets scales with the uniform magnetic susceptibility $\chi$ in 3D and $1/T_{1}T \propto \chi^{3/2}$ in 2D.~\cite{Moriya-1985,Hatatani-1995} The exponents have been calculated at the FM QCP and the expected value is $n = 4/3$ in 3D, while in 2D there is no long range FM order and the SCR theory predicts $1/T_{1}T \propto T^{-3/2}(-\ln T)^{-3/2} \approx T^{-1.4}$ for $T \rightarrow 0$.~\cite{{RevModPhys.73.797},{Hatatani-1995}} Our exponents are therefore not close to these predictions, they are smaller.  In addition, the unusual power law exponent in the thermodynamic quantities support our present findings, and suggests the inadequacy of the itinerant spin-fluctuations theory for the FM fluctuation in the close proximity of the QCP.~\cite{Steppke-2013}
 
Previous nuclear magnetic resonance (NMR) experiments have detected a spin-lattice relaxation rate $1/T_1T$ which follows a $\sim T^{-n}$ function also with a small $n = 0.75$ in a broad temperature range above 2\,K.~\cite{Sarkar-2012} Although NMR experiments were performed in an external magnetic field and crystalline electric field effects have to be considered in this temperature range, the exponent is still small compared to other Yb-based FM systems studied at the FM QPT in the same $T$ range.~\cite{Carretta-2009}

\begin{figure}[t]
\includegraphics[scale=0.5, trim= 40 120 40 40]{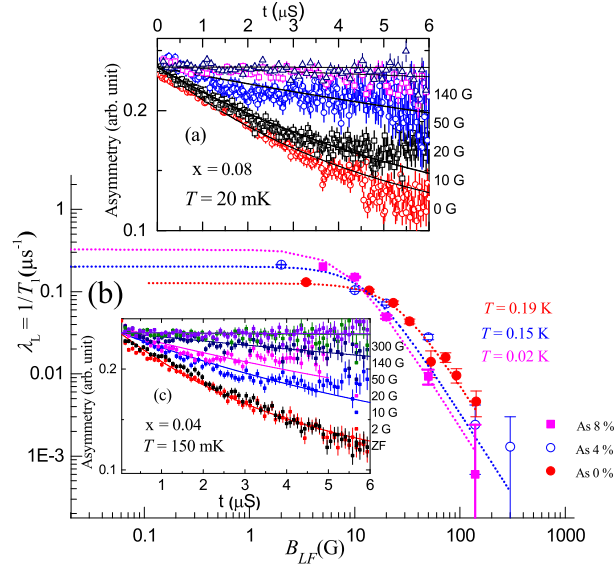}
\caption{\label{fig:fig4} The main panel shows the magnetic-field dependence of the dynamic muon-spin relaxation rate $\lambda_{L}$ for $x = 0$, 0.04 and 0.08. The inset shows the magnetic-field dependence of the muon-spin polarization at $T = 150$\,mK for $x = 0.04$. The top panel shows the magnetic-field dependence of the muon-spin polarization at $T = 20$\,mK for $x = 0.08$. The dotted line represents Redfield fits (see main text).}
\end{figure}

To understand the nature of magnetic correlations, in particular whether the magnetism is static or dynamic, $\mu$SR experiments in different longitudinal fields (LFs) up to $B_{\mathrm{LF}} = 300$\,G were performed. We found that above \TC\ and with small applied fields the muon spin-relaxation is completely suppressed in the samples with $x = 0.0$ and 0.04 (see Ref.~\onlinecite{Spehling-2012} and Fig.~\ref{fig:fig4}c). This indicates that the zero-field relaxation in the magnetic ordered state originates from static internal fields on the microsecond timescale.
Above the ordering temperature, we could fit all data for all As concentrations with a simple exponential function $P(t) = P(0)e^{-\lambda_{L} t}$ as shown in Fig.~\ref{fig:fig4} (solid lines). The single exponential fit in the LF suggests that the system is free from magnetic disorder and is intrinsically homogeneous. The spin dynamics is characterized by a narrow distribution of correlation times for $x = 0$ as well for $x = 0.08$ near the FM QCP. With increasing field, the relaxation rates $\lambda_{L}$ decreases. The field dependence of $\lambda_L$ is presented in the main panel of Fig.~\ref{fig:fig4}b for all $x\leq 0.1$. They show very similar values. From the magnetic field dependence of $\lambda_L$, the spin autocorrelation time, $\tau_c$ can be estimated using $\lambda_L(B_{LF})=(2\gamma^{2}_{\mu}\left\langle B^{2}_{\mathrm{loc}}\right\rangle\tau_c)/\left[ 1+(\gamma^{2}_{\mu}B^{2}_{LF}\tau^{2}_{c})\right]$. This is so called the Redfield formalism where $\tau_c$ is independent of the applied magnetic field $B_{LF}$. Here, $B_\mathrm{loc}(t)$ describes the time-varying local magnetic field at the muon site due to fluctuations of neighboring Yb $4f$ moments, with a local time averaged second moment $\Delta^2=\gamma^{2}_{\mu}\left\langle B^{2}_{\mathrm{loc}}\right\rangle$, and a single fluctuation time $\tau_c$, meaning a single exponential auto-correlation function characterized by a single time constant. For $\hbar\omega$\,$<<$\,$k_BT$ ($\omega$ giving the spin fluctuation frequency), the fluctuation-dissipation theorem \cite{Toll-dissipation-theorem} relates $\tau_c$ to the imaginary component of the local $q$-independent $f$-electron dynamic susceptibility, i.e., $\tau_c(B)=(k_BT)[\chi''/\omega]$. Dotted curves in the main panel of Fig.~\ref{fig:fig4}b are fits to the experimental data. The results of the fits give $\Delta^2 \approx 0.1$, 0.1, and 0.108\,(MHz) and $\tau_c \approx 6.00\times 10^{-7}$, 9.11$\times 10^{-7}$, and 14.11$\times 10^{-7}$\,s for $x = 0.00$, 0.04, and 0.08, respectively. With increasing $x$, $\tau_c$ becomes larger and it becomes even larger for $x \rightarrow$ \xc. The estimated $\tau_c$ value for $x = 0.08$ is much larger than that found in YbRh$_2$Si$_2$~\cite{Ishida-musr-YbRh2Si2} and in CeFePO~\cite{Lausberg-CeFePO-2012}, respectively. These fluctuations can therefore be considered as very slow critical spin fluctuations. On the other hand similar $\Delta^2$ values for all the concentration suggest rather homogeneous system because $\Delta^2$ is the second moment of the field distribution at the muon stopping sites. If the system is inhomogeneous then one would expect rather different field distribution for different As concentration.
\begin{table*}
\caption{\label{tab:table1} Comparison of results obtained for YbNi$_4$(P$_{1-x}$As$_x$)$_2$ with
different As concentration.}
\begin{ruledtabular}
\begin{tabular}{ccccc}
 $x$ & $n^{\prime}$ & f$_{\mu}$(20 mK) & $B_{\mathrm{local}}$(20 mK) & $\mu_{\mathrm{ord}}$ (20 mK in $\mu_B$) \\ \hline
 0.00 & 0.21\,$\pm$ 0.02 & 0.1993 & 13.87 G & 0.04\\
 0.04 & 0.28\,$\pm$ 0.05 & 0.0989 & 6.64 G & 0.021\\
 0.08 & -   & - & -& $< 0.005$\\
\end{tabular}
\end{ruledtabular}
\end{table*}
\\
\indent
In conclusion, the present $\mu$SR experiments show that by introducing a small amount of As at the P site in the system \YNPA\, the magnetic order remains homogeneous and the static ordered Yb moment is suppressed continuously from about 0.04\,\mub\ to values lower than 0.005\,\mub\ for $x \rightarrow$ \xc\ = 0.1. There is no evidence for disorder effect in the $\mu$SR samples. Notably, in the paramagnetic region of the phase diagram and across the FM QCP the dynamic muon-spin relaxation rate 1/$T_{1}T \propto T^{-n}$ with $n \approx 1$ indicating 1/$T_1$ is approximately $T$ independent. The critical fluctuations are very slow in the pure compound and become even slower when approaching the FM QCP. All these findings support the presence of a clean FM QCP in this system. However, the power laws observed in the $T$-dependence of thermodynamic quantities and in $1/T_{1}T$ remain to be understood.

We thank Z. Huesges for having prepared the samples for $\mu$SR experiments and C. Geibel for fruitful discussions. We are thankful to DFG for the financial assistance through the grant no. SA 2426/1-1, KR 3831/4-1, BR 4110/1-1, and SFB 1143 for the project C02.
\bibliography{ybnip}

\begin{thebibliography}{46}%
\makeatletter
\providecommand \@ifxundefined [1]{%
 \@ifx{#1\undefined}
}%
\providecommand \@ifnum [1]{%
 \ifnum #1\expandafter \@firstoftwo
 \else \expandafter \@secondoftwo
 \fi
}%
\providecommand \@ifx [1]{%
 \ifx #1\expandafter \@firstoftwo
 \else \expandafter \@secondoftwo
 \fi
}%
\providecommand \natexlab [1]{#1}%
\providecommand \enquote  [1]{``#1''}%
\providecommand \bibnamefont  [1]{#1}%
\providecommand \bibfnamefont [1]{#1}%
\providecommand \citenamefont [1]{#1}%
\providecommand \href@noop [0]{\@secondoftwo}%
\providecommand \href [0]{\begingroup \@sanitize@url \@href}%
\providecommand \@href[1]{\@@startlink{#1}\@@href}%
\providecommand \@@href[1]{\endgroup#1\@@endlink}%
\providecommand \@sanitize@url [0]{\catcode `\\12\catcode `\$12\catcode
  `\&12\catcode `\#12\catcode `\^12\catcode `\_12\catcode `\%12\relax}%
\providecommand \@@startlink[1]{}%
\providecommand \@@endlink[0]{}%
\providecommand \url  [0]{\begingroup\@sanitize@url \@url }%
\providecommand \@url [1]{\endgroup\@href {#1}{\urlprefix }}%
\providecommand \urlprefix  [0]{URL }%
\providecommand \Eprint [0]{\href }%
\providecommand \doibase [0]{http://dx.doi.org/}%
\providecommand \selectlanguage [0]{\@gobble}%
\providecommand \bibinfo  [0]{\@secondoftwo}%
\providecommand \bibfield  [0]{\@secondoftwo}%
\providecommand \translation [1]{[#1]}%
\providecommand \BibitemOpen [0]{}%
\providecommand \bibitemStop [0]{}%
\providecommand \bibitemNoStop [0]{.\EOS\space}%
\providecommand \EOS [0]{\spacefactor3000\relax}%
\providecommand \BibitemShut  [1]{\csname bibitem#1\endcsname}%
\let\auto@bib@innerbib\@empty
\bibitem [{\citenamefont {Canfield}(2008)}]{Focus-QPT-2008-1}%
  \BibitemOpen
  \bibfield  {author} {\bibinfo {author} {\bibfnamefont {P.~C.}\ \bibnamefont
  {Canfield}},\ }\href@noop {} {\bibfield  {journal} {\bibinfo  {journal} {Nat.
  Phys.}\ }\textbf {\bibinfo {volume} {4}},\ \bibinfo {pages} {167} (\bibinfo
  {year} {2008})}\BibitemShut {NoStop}%
\bibitem [{\citenamefont {Broun}(2008)}]{Focus-QPT-2008-2}%
  \BibitemOpen
  \bibfield  {author} {\bibinfo {author} {\bibfnamefont {D.~M.}\ \bibnamefont
  {Broun}},\ }\href@noop {} {\bibfield  {journal} {\bibinfo  {journal} {Nat.
  Phys.}\ }\textbf {\bibinfo {volume} {4}},\ \bibinfo {pages} {170} (\bibinfo
  {year} {2008})}\BibitemShut {NoStop}%
\bibitem [{\citenamefont {Sachdev}(2008)}]{Focus-QPT-2008-3}%
  \BibitemOpen
  \bibfield  {author} {\bibinfo {author} {\bibfnamefont {S.}~\bibnamefont
  {Sachdev}},\ }\href@noop {} {\bibfield  {journal} {\bibinfo  {journal} {Nat.
  Phys.}\ }\textbf {\bibinfo {volume} {4}},\ \bibinfo {pages} {173} (\bibinfo
  {year} {2008})}\BibitemShut {NoStop}%
\bibitem [{\citenamefont {Gegenwart}\ \emph {et~al.}(2008)\citenamefont
  {Gegenwart}, \citenamefont {Si},\ and\ \citenamefont
  {Steglich}}]{Focus-QPT-2008-4}%
  \BibitemOpen
  \bibfield  {author} {\bibinfo {author} {\bibfnamefont {P.}~\bibnamefont
  {Gegenwart}}, \bibinfo {author} {\bibfnamefont {Q.}~\bibnamefont {Si}}, \
  and\ \bibinfo {author} {\bibfnamefont {F.}~\bibnamefont {Steglich}},\
  }\href@noop {} {\bibfield  {journal} {\bibinfo  {journal} {Nat. Phys.}\
  }\textbf {\bibinfo {volume} {4}},\ \bibinfo {pages} {186} (\bibinfo {year}
  {2008})}\BibitemShut {NoStop}%
\bibitem [{\citenamefont {Brando}\ \emph {et~al.}(2016)\citenamefont {Brando},
  \citenamefont {Belitz}, \citenamefont {Grosche},\ and\ \citenamefont
  {Kirkpatrick}}]{Brando-2015}%
  \BibitemOpen
  \bibfield  {author} {\bibinfo {author} {\bibfnamefont {M.}~\bibnamefont
  {Brando}}, \bibinfo {author} {\bibfnamefont {D.}~\bibnamefont {Belitz}},
  \bibinfo {author} {\bibfnamefont {F.~M.}\ \bibnamefont {Grosche}}, \ and\
  \bibinfo {author} {\bibfnamefont {T.~R.}\ \bibnamefont {Kirkpatrick}},\
  }\href {\doibase 10.1103/RevModPhys.88.039901} {\bibfield  {journal}
  {\bibinfo  {journal} {Rev. Mod. Phys.}\ }\textbf {\bibinfo {volume} {88}},\
  \bibinfo {pages} {039901} (\bibinfo {year} {2016})}\BibitemShut {NoStop}%
\bibitem [{\citenamefont {Belitz}\ \emph {et~al.}(1999)\citenamefont {Belitz},
  \citenamefont {Kirkpatrick},\ and\ \citenamefont {Vojta}}]{Belitz-1999}%
  \BibitemOpen
  \bibfield  {author} {\bibinfo {author} {\bibfnamefont {D.}~\bibnamefont
  {Belitz}}, \bibinfo {author} {\bibfnamefont {T.~R.}\ \bibnamefont
  {Kirkpatrick}}, \ and\ \bibinfo {author} {\bibfnamefont {T.}~\bibnamefont
  {Vojta}},\ }\href {\doibase 10.1103/PhysRevLett.82.4707} {\bibfield
  {journal} {\bibinfo  {journal} {Phys. Rev. Lett.}\ }\textbf {\bibinfo
  {volume} {82}},\ \bibinfo {pages} {4707} (\bibinfo {year}
  {1999})}\BibitemShut {NoStop}%
\bibitem [{\citenamefont {Chubukov}\ \emph {et~al.}(2004)\citenamefont
  {Chubukov}, \citenamefont {P\'epin},\ and\ \citenamefont
  {Rech}}]{Chubukov-2004}%
  \BibitemOpen
  \bibfield  {author} {\bibinfo {author} {\bibfnamefont {A.~V.}\ \bibnamefont
  {Chubukov}}, \bibinfo {author} {\bibfnamefont {C.}~\bibnamefont {P\'epin}}, \
  and\ \bibinfo {author} {\bibfnamefont {J.}~\bibnamefont {Rech}},\ }\href
  {\doibase 10.1103/PhysRevLett.92.147003} {\bibfield  {journal} {\bibinfo
  {journal} {Phys. Rev. Lett.}\ }\textbf {\bibinfo {volume} {92}},\ \bibinfo
  {pages} {147003} (\bibinfo {year} {2004})}\BibitemShut {NoStop}%
\bibitem [{\citenamefont {Pfleiderer}\ \emph {et~al.}(1997)\citenamefont
  {Pfleiderer}, \citenamefont {McMullan}, \citenamefont {Julian},\ and\
  \citenamefont {Lonzarich}}]{Pfleiderer-1997}%
  \BibitemOpen
  \bibfield  {author} {\bibinfo {author} {\bibfnamefont {C.}~\bibnamefont
  {Pfleiderer}}, \bibinfo {author} {\bibfnamefont {G.~J.}\ \bibnamefont
  {McMullan}}, \bibinfo {author} {\bibfnamefont {S.~R.}\ \bibnamefont
  {Julian}}, \ and\ \bibinfo {author} {\bibfnamefont {G.~G.}\ \bibnamefont
  {Lonzarich}},\ }\href@noop {} {\bibfield  {journal} {\bibinfo  {journal}
  {Phys. Rev. B}\ }\textbf {\bibinfo {volume} {55}},\ \bibinfo {pages} {8330}
  (\bibinfo {year} {1997})}\BibitemShut {NoStop}%
\bibitem [{\citenamefont {Uhlarz}\ \emph {et~al.}(2004)\citenamefont {Uhlarz},
  \citenamefont {Pfleiderer},\ and\ \citenamefont {Hayden}}]{Uhlarz-2004}%
  \BibitemOpen
  \bibfield  {author} {\bibinfo {author} {\bibfnamefont {M.}~\bibnamefont
  {Uhlarz}}, \bibinfo {author} {\bibfnamefont {C.}~\bibnamefont {Pfleiderer}},
  \ and\ \bibinfo {author} {\bibfnamefont {S.~M.}\ \bibnamefont {Hayden}},\
  }\href@noop {} {\bibfield  {journal} {\bibinfo  {journal} {Phys. Rev. Lett.}\
  }\textbf {\bibinfo {volume} {93}},\ \bibinfo {pages} {256404} (\bibinfo
  {year} {2004})}\BibitemShut {NoStop}%
\bibitem [{\citenamefont {Aoki}\ \emph {et~al.}(2011)\citenamefont {Aoki},
  \citenamefont {Combier}, \citenamefont {Taufour}, \citenamefont {Matsuda},
  \citenamefont {Knebel}, \citenamefont {Kotegawa},\ and\ \citenamefont
  {Flouquet}}]{Aoki-2011}%
  \BibitemOpen
  \bibfield  {author} {\bibinfo {author} {\bibfnamefont {D.}~\bibnamefont
  {Aoki}}, \bibinfo {author} {\bibfnamefont {T.}~\bibnamefont {Combier}},
  \bibinfo {author} {\bibfnamefont {V.}~\bibnamefont {Taufour}}, \bibinfo
  {author} {\bibfnamefont {T.~D.}\ \bibnamefont {Matsuda}}, \bibinfo {author}
  {\bibfnamefont {G.}~\bibnamefont {Knebel}}, \bibinfo {author} {\bibfnamefont
  {H.}~\bibnamefont {Kotegawa}}, \ and\ \bibinfo {author} {\bibfnamefont
  {J.}~\bibnamefont {Flouquet}},\ }\href@noop {} {\bibfield  {journal}
  {\bibinfo  {journal} {J. Phys. Soc. Jpn.}\ }\textbf {\bibinfo {volume}
  {80}},\ \bibinfo {pages} {094711} (\bibinfo {year} {2011})}\BibitemShut
  {NoStop}%
\bibitem [{\citenamefont {Belitz}\ and\ \citenamefont
  {Kirkpatrick}(1997)}]{Belitz-1997}%
  \BibitemOpen
  \bibfield  {author} {\bibinfo {author} {\bibfnamefont {D.}~\bibnamefont
  {Belitz}}\ and\ \bibinfo {author} {\bibfnamefont {T.~R.}\ \bibnamefont
  {Kirkpatrick}},\ }\href@noop {} {\bibfield  {journal} {\bibinfo  {journal}
  {Phys. Rev. B}\ }\textbf {\bibinfo {volume} {56}},\ \bibinfo {pages} {6513}
  (\bibinfo {year} {1997})}\BibitemShut {NoStop}%
\bibitem [{\citenamefont {Conduit}\ \emph {et~al.}(2009)\citenamefont
  {Conduit}, \citenamefont {Green},\ and\ \citenamefont
  {Simons}}]{Conduit-2009}%
  \BibitemOpen
  \bibfield  {author} {\bibinfo {author} {\bibfnamefont {G.~J.}\ \bibnamefont
  {Conduit}}, \bibinfo {author} {\bibfnamefont {A.~G.}\ \bibnamefont {Green}},
  \ and\ \bibinfo {author} {\bibfnamefont {B.~D.}\ \bibnamefont {Simons}},\
  }\href {\doibase 10.1103/PhysRevLett.103.207201} {\bibfield  {journal}
  {\bibinfo  {journal} {Phys. Rev. Lett.}\ }\textbf {\bibinfo {volume} {103}},\
  \bibinfo {pages} {207201} (\bibinfo {year} {2009})}\BibitemShut {NoStop}%
\bibitem [{\citenamefont {Karahasanovic}\ \emph {et~al.}(2012)\citenamefont
  {Karahasanovic}, \citenamefont {Kr{\"u}ger},\ and\ \citenamefont
  {Green}}]{Karahasanovic-2012}%
  \BibitemOpen
  \bibfield  {author} {\bibinfo {author} {\bibfnamefont {U.}~\bibnamefont
  {Karahasanovic}}, \bibinfo {author} {\bibfnamefont {F.}~\bibnamefont
  {Kr{\"u}ger}}, \ and\ \bibinfo {author} {\bibfnamefont {A.~G.}\ \bibnamefont
  {Green}},\ }\href@noop {} {\bibfield  {journal} {\bibinfo  {journal} {Phys.
  Rev. B}\ }\textbf {\bibinfo {volume} {85}},\ \bibinfo {pages} {165111}
  (\bibinfo {year} {2012})}\BibitemShut {NoStop}%
\bibitem [{\citenamefont {Brando}\ \emph {et~al.}(2008)\citenamefont {Brando},
  \citenamefont {Duncan}, \citenamefont {Moroni-Klementowicz}, \citenamefont
  {Albrecht}, \citenamefont {Gr\"uner}, \citenamefont {Ballou},\ and\
  \citenamefont {Grosche}}]{Brando-2008}%
  \BibitemOpen
  \bibfield  {author} {\bibinfo {author} {\bibfnamefont {M.}~\bibnamefont
  {Brando}}, \bibinfo {author} {\bibfnamefont {W.~J.}\ \bibnamefont {Duncan}},
  \bibinfo {author} {\bibfnamefont {D.}~\bibnamefont {Moroni-Klementowicz}},
  \bibinfo {author} {\bibfnamefont {C.}~\bibnamefont {Albrecht}}, \bibinfo
  {author} {\bibfnamefont {D.}~\bibnamefont {Gr\"uner}}, \bibinfo {author}
  {\bibfnamefont {R.}~\bibnamefont {Ballou}}, \ and\ \bibinfo {author}
  {\bibfnamefont {F.~M.}\ \bibnamefont {Grosche}},\ }\href {\doibase
  10.1103/PhysRevLett.101.026401} {\bibfield  {journal} {\bibinfo  {journal}
  {Phys. Rev. Lett.}\ }\textbf {\bibinfo {volume} {101}},\ \bibinfo {pages}
  {026401} (\bibinfo {year} {2008})}\BibitemShut {NoStop}%
\bibitem [{\citenamefont {Kotegawa}\ \emph {et~al.}(2013)\citenamefont
  {Kotegawa}, \citenamefont {Toyama}, \citenamefont {Kitagawa}, \citenamefont
  {Tou}, \citenamefont {Yamauchi}, \citenamefont {Matsuoka},\ and\
  \citenamefont {Sugawara}}]{Kotegawa-2013}%
  \BibitemOpen
  \bibfield  {author} {\bibinfo {author} {\bibfnamefont {H.}~\bibnamefont
  {Kotegawa}}, \bibinfo {author} {\bibfnamefont {T.}~\bibnamefont {Toyama}},
  \bibinfo {author} {\bibfnamefont {S.}~\bibnamefont {Kitagawa}}, \bibinfo
  {author} {\bibfnamefont {H.}~\bibnamefont {Tou}}, \bibinfo {author}
  {\bibfnamefont {R.}~\bibnamefont {Yamauchi}}, \bibinfo {author}
  {\bibfnamefont {E.}~\bibnamefont {Matsuoka}}, \ and\ \bibinfo {author}
  {\bibfnamefont {H.}~\bibnamefont {Sugawara}},\ }\href@noop {} {\bibfield
  {journal} {\bibinfo  {journal} {J. Phys. Soc. Jpn.}\ }\textbf {\bibinfo
  {volume} {82}},\ \bibinfo {pages} {123711} (\bibinfo {year}
  {2013})}\BibitemShut {NoStop}%
\bibitem [{\citenamefont {Westerkamp}\ \emph {et~al.}(2009)\citenamefont
  {Westerkamp}, \citenamefont {Deppe}, \citenamefont {Kuchler}, \citenamefont
  {Brando}, \citenamefont {Geibel}, \citenamefont {Gegenwart}, \citenamefont
  {Pikul},\ and\ \citenamefont {Steglich}}]{Westerkamp-2009}%
  \BibitemOpen
  \bibfield  {author} {\bibinfo {author} {\bibfnamefont {T.}~\bibnamefont
  {Westerkamp}}, \bibinfo {author} {\bibfnamefont {M.}~\bibnamefont {Deppe}},
  \bibinfo {author} {\bibfnamefont {R.}~\bibnamefont {Kuchler}}, \bibinfo
  {author} {\bibfnamefont {M.}~\bibnamefont {Brando}}, \bibinfo {author}
  {\bibfnamefont {C.}~\bibnamefont {Geibel}}, \bibinfo {author} {\bibfnamefont
  {P.}~\bibnamefont {Gegenwart}}, \bibinfo {author} {\bibfnamefont {A.~P.}\
  \bibnamefont {Pikul}}, \ and\ \bibinfo {author} {\bibfnamefont
  {F.}~\bibnamefont {Steglich}},\ }\href@noop {} {\bibfield  {journal}
  {\bibinfo  {journal} {Phys. Rev. Lett.}\ }\textbf {\bibinfo {volume} {102}},\
  \bibinfo {pages} {206404} (\bibinfo {year} {2009})}\BibitemShut {NoStop}%
\bibitem [{\citenamefont {Ubaid-Kassis}\ \emph {et~al.}(2010)\citenamefont
  {Ubaid-Kassis}, \citenamefont {Vojta},\ and\ \citenamefont
  {Schroeder}}]{Ubaid-Kassis-2010}%
  \BibitemOpen
  \bibfield  {author} {\bibinfo {author} {\bibfnamefont {S.}~\bibnamefont
  {Ubaid-Kassis}}, \bibinfo {author} {\bibfnamefont {T.}~\bibnamefont {Vojta}},
  \ and\ \bibinfo {author} {\bibfnamefont {A.}~\bibnamefont {Schroeder}},\
  }\href@noop {} {\bibfield  {journal} {\bibinfo  {journal} {Phys. Rev. Lett.}\
  }\textbf {\bibinfo {volume} {104}},\ \bibinfo {pages} {066402} (\bibinfo
  {year} {2010})}\BibitemShut {NoStop}%
\bibitem [{\citenamefont {Lausberg}\ \emph {et~al.}(2012)\citenamefont
  {Lausberg}, \citenamefont {Spehling}, \citenamefont {Steppke}, \citenamefont
  {Jesche}, \citenamefont {Luetkens}, \citenamefont {Amato}, \citenamefont
  {Baines}, \citenamefont {Krellner}, \citenamefont {Brando}, \citenamefont
  {Geibel}, \citenamefont {Klauss},\ and\ \citenamefont
  {Steglich}}]{Lausberg-CeFePO-2012}%
  \BibitemOpen
  \bibfield  {author} {\bibinfo {author} {\bibfnamefont {S.}~\bibnamefont
  {Lausberg}}, \bibinfo {author} {\bibfnamefont {J.}~\bibnamefont {Spehling}},
  \bibinfo {author} {\bibfnamefont {A.}~\bibnamefont {Steppke}}, \bibinfo
  {author} {\bibfnamefont {A.}~\bibnamefont {Jesche}}, \bibinfo {author}
  {\bibfnamefont {H.}~\bibnamefont {Luetkens}}, \bibinfo {author}
  {\bibfnamefont {A.}~\bibnamefont {Amato}}, \bibinfo {author} {\bibfnamefont
  {C.}~\bibnamefont {Baines}}, \bibinfo {author} {\bibfnamefont
  {C.}~\bibnamefont {Krellner}}, \bibinfo {author} {\bibfnamefont
  {M.}~\bibnamefont {Brando}}, \bibinfo {author} {\bibfnamefont
  {C.}~\bibnamefont {Geibel}}, \bibinfo {author} {\bibfnamefont {H.-H.}\
  \bibnamefont {Klauss}}, \ and\ \bibinfo {author} {\bibfnamefont
  {F.}~\bibnamefont {Steglich}},\ }\href {\doibase
  10.1103/PhysRevLett.109.216402} {\bibfield  {journal} {\bibinfo  {journal}
  {Phys. Rev. Lett.}\ }\textbf {\bibinfo {volume} {109}},\ \bibinfo {pages}
  {216402} (\bibinfo {year} {2012})}\BibitemShut {NoStop}%
\bibitem [{\citenamefont {Taufour}\ \emph {et~al.}(2010)\citenamefont
  {Taufour}, \citenamefont {Aoki}, \citenamefont {Knebel},\ and\ \citenamefont
  {Flouquet}}]{Taufour-2010}%
  \BibitemOpen
  \bibfield  {author} {\bibinfo {author} {\bibfnamefont {V.}~\bibnamefont
  {Taufour}}, \bibinfo {author} {\bibfnamefont {D.}~\bibnamefont {Aoki}},
  \bibinfo {author} {\bibfnamefont {G.}~\bibnamefont {Knebel}}, \ and\ \bibinfo
  {author} {\bibfnamefont {J.}~\bibnamefont {Flouquet}},\ }\href {\doibase
  10.1103/PhysRevLett.105.217201} {\bibfield  {journal} {\bibinfo  {journal}
  {Phys. Rev. Lett.}\ }\textbf {\bibinfo {volume} {105}},\ \bibinfo {pages}
  {217201} (\bibinfo {year} {2010})}\BibitemShut {NoStop}%
\bibitem [{\citenamefont {Huy}\ \emph {et~al.}(2007)\citenamefont {Huy},
  \citenamefont {Gasparini}, \citenamefont {de~Nijs}, \citenamefont {Huang},
  \citenamefont {Klaasse}, \citenamefont {Gortenmulder}, \citenamefont
  {de~Visser}, \citenamefont {Hamann}, \citenamefont {G\"orlach},\ and\
  \citenamefont {L\"ohneysen}}]{Huy-2007}%
  \BibitemOpen
  \bibfield  {author} {\bibinfo {author} {\bibfnamefont {N.~T.}\ \bibnamefont
  {Huy}}, \bibinfo {author} {\bibfnamefont {A.}~\bibnamefont {Gasparini}},
  \bibinfo {author} {\bibfnamefont {D.~E.}\ \bibnamefont {de~Nijs}}, \bibinfo
  {author} {\bibfnamefont {Y.}~\bibnamefont {Huang}}, \bibinfo {author}
  {\bibfnamefont {J.~C.~P.}\ \bibnamefont {Klaasse}}, \bibinfo {author}
  {\bibfnamefont {T.}~\bibnamefont {Gortenmulder}}, \bibinfo {author}
  {\bibfnamefont {A.}~\bibnamefont {de~Visser}}, \bibinfo {author}
  {\bibfnamefont {A.}~\bibnamefont {Hamann}}, \bibinfo {author} {\bibfnamefont
  {T.}~\bibnamefont {G\"orlach}}, \ and\ \bibinfo {author} {\bibfnamefont
  {H.~v.}\ \bibnamefont {L\"ohneysen}},\ }\href {\doibase
  10.1103/PhysRevLett.99.067006} {\bibfield  {journal} {\bibinfo  {journal}
  {Phys. Rev. Lett.}\ }\textbf {\bibinfo {volume} {99}},\ \bibinfo {pages}
  {067006} (\bibinfo {year} {2007})}\BibitemShut {NoStop}%
\bibitem [{\citenamefont {Steppke}\ \emph {et~al.}(2013)\citenamefont
  {Steppke}, \citenamefont {Kuechler}, \citenamefont {Lausberg}, \citenamefont
  {Lengyel}, \citenamefont {Steinke}, \citenamefont {Borth}, \citenamefont
  {Luehmann}, \citenamefont {Krellner}, \citenamefont {Nicklas}, \citenamefont
  {Geibel}, \citenamefont {Steglich},\ and\ \citenamefont
  {Brando}}]{Steppke-2013}%
  \BibitemOpen
  \bibfield  {author} {\bibinfo {author} {\bibfnamefont {A.}~\bibnamefont
  {Steppke}}, \bibinfo {author} {\bibfnamefont {R.}~\bibnamefont {Kuechler}},
  \bibinfo {author} {\bibfnamefont {S.}~\bibnamefont {Lausberg}}, \bibinfo
  {author} {\bibfnamefont {E.}~\bibnamefont {Lengyel}}, \bibinfo {author}
  {\bibfnamefont {L.}~\bibnamefont {Steinke}}, \bibinfo {author} {\bibfnamefont
  {R.}~\bibnamefont {Borth}}, \bibinfo {author} {\bibfnamefont
  {T.}~\bibnamefont {Luehmann}}, \bibinfo {author} {\bibfnamefont
  {C.}~\bibnamefont {Krellner}}, \bibinfo {author} {\bibfnamefont
  {M.}~\bibnamefont {Nicklas}}, \bibinfo {author} {\bibfnamefont
  {C.}~\bibnamefont {Geibel}}, \bibinfo {author} {\bibfnamefont
  {F.}~\bibnamefont {Steglich}}, \ and\ \bibinfo {author} {\bibfnamefont
  {M.}~\bibnamefont {Brando}},\ }\href {\doibase 10.1126/science.1230583}
  {\bibfield  {journal} {\bibinfo  {journal} {Science}\ }\textbf {\bibinfo
  {volume} {339}},\ \bibinfo {pages} {933} (\bibinfo {year}
  {2013})}\BibitemShut {NoStop}%
\bibitem [{\citenamefont {Zhu}\ \emph {et~al.}(2003)\citenamefont {Zhu},
  \citenamefont {Garst}, \citenamefont {Rosch},\ and\ \citenamefont
  {Si}}]{Zhu-2003}%
  \BibitemOpen
  \bibfield  {author} {\bibinfo {author} {\bibfnamefont {L.}~\bibnamefont
  {Zhu}}, \bibinfo {author} {\bibfnamefont {M.}~\bibnamefont {Garst}}, \bibinfo
  {author} {\bibfnamefont {A.}~\bibnamefont {Rosch}}, \ and\ \bibinfo {author}
  {\bibfnamefont {Q.}~\bibnamefont {Si}},\ }\href@noop {} {\bibfield  {journal}
  {\bibinfo  {journal} {Phys. Rev. Lett.}\ }\textbf {\bibinfo {volume} {91}},\
  \bibinfo {pages} {066404} (\bibinfo {year} {2003})}\BibitemShut {NoStop}%
\bibitem [{\citenamefont {Krellner}\ \emph {et~al.}(2011)\citenamefont
  {Krellner}, \citenamefont {Lausberg}, \citenamefont {Steppke}, \citenamefont
  {Brando}, \citenamefont {Pedrero}, \citenamefont {Pfau}, \citenamefont
  {Tencé}, \citenamefont {Rosner}, \citenamefont {Steglich},\ and\
  \citenamefont {Geibel}}]{Krellner-2011}%
  \BibitemOpen
  \bibfield  {author} {\bibinfo {author} {\bibfnamefont {C.}~\bibnamefont
  {Krellner}}, \bibinfo {author} {\bibfnamefont {S.}~\bibnamefont {Lausberg}},
  \bibinfo {author} {\bibfnamefont {A.}~\bibnamefont {Steppke}}, \bibinfo
  {author} {\bibfnamefont {M.}~\bibnamefont {Brando}}, \bibinfo {author}
  {\bibfnamefont {L.}~\bibnamefont {Pedrero}}, \bibinfo {author} {\bibfnamefont
  {H.}~\bibnamefont {Pfau}}, \bibinfo {author} {\bibfnamefont {S.}~\bibnamefont
  {Tencé}}, \bibinfo {author} {\bibfnamefont {H.}~\bibnamefont {Rosner}},
  \bibinfo {author} {\bibfnamefont {F.}~\bibnamefont {Steglich}}, \ and\
  \bibinfo {author} {\bibfnamefont {C.}~\bibnamefont {Geibel}},\ }\href
  {http://stacks.iop.org/1367-2630/13/i=10/a=103014} {\bibfield  {journal}
  {\bibinfo  {journal} {New Journal of Physics}\ }\textbf {\bibinfo {volume}
  {13}},\ \bibinfo {pages} {103014} (\bibinfo {year} {2011})}\BibitemShut
  {NoStop}%
\bibitem [{\citenamefont {Spehling}\ \emph {et~al.}(2012)\citenamefont
  {Spehling}, \citenamefont {G\"unther}, \citenamefont {Krellner},
  \citenamefont {Y\`eche}, \citenamefont {Luetkens}, \citenamefont {Baines},
  \citenamefont {Geibel},\ and\ \citenamefont {Klauss}}]{Spehling-2012}%
  \BibitemOpen
  \bibfield  {author} {\bibinfo {author} {\bibfnamefont {J.}~\bibnamefont
  {Spehling}}, \bibinfo {author} {\bibfnamefont {M.}~\bibnamefont {G\"unther}},
  \bibinfo {author} {\bibfnamefont {C.}~\bibnamefont {Krellner}}, \bibinfo
  {author} {\bibfnamefont {N.}~\bibnamefont {Y\`eche}}, \bibinfo {author}
  {\bibfnamefont {H.}~\bibnamefont {Luetkens}}, \bibinfo {author}
  {\bibfnamefont {C.}~\bibnamefont {Baines}}, \bibinfo {author} {\bibfnamefont
  {C.}~\bibnamefont {Geibel}}, \ and\ \bibinfo {author} {\bibfnamefont {H.-H.}\
  \bibnamefont {Klauss}},\ }\href@noop {} {\bibfield  {journal} {\bibinfo
  {journal} {Phys. Rev. B}\ }\textbf {\bibinfo {volume} {85}},\ \bibinfo
  {pages} {140406} (\bibinfo {year} {2012})}\BibitemShut {NoStop}%
\bibitem [{\citenamefont {Pivan}\ \emph {et~al.}(1989)\citenamefont {Pivan},
  \citenamefont {Guerin}, \citenamefont {Ghadraoui},\ and\ \citenamefont
  {Rafiq}}]{PIVAN1989285}%
  \BibitemOpen
  \bibfield  {author} {\bibinfo {author} {\bibfnamefont {J.}~\bibnamefont
  {Pivan}}, \bibinfo {author} {\bibfnamefont {R.}~\bibnamefont {Guerin}},
  \bibinfo {author} {\bibfnamefont {E.~E.}\ \bibnamefont {Ghadraoui}}, \ and\
  \bibinfo {author} {\bibfnamefont {M.}~\bibnamefont {Rafiq}},\ }\href
  {\doibase http://dx.doi.org/10.1016/0022-5088(89)90123-9} {\bibfield
  {journal} {\bibinfo  {journal} {Journal of the Less Common Metals}\ }\textbf
  {\bibinfo {volume} {153}},\ \bibinfo {pages} {285 } (\bibinfo {year}
  {1989})}\BibitemShut {NoStop}%
\bibitem [{\citenamefont {Krellner}\ and\ \citenamefont
  {Geibel}(2012)}]{Krellner-jpcm-2012}%
  \BibitemOpen
  \bibfield  {author} {\bibinfo {author} {\bibfnamefont {C.}~\bibnamefont
  {Krellner}}\ and\ \bibinfo {author} {\bibfnamefont {C.}~\bibnamefont
  {Geibel}},\ }\href {http://stacks.iop.org/1742-6596/391/i=1/a=012032}
  {\bibfield  {journal} {\bibinfo  {journal} {Journal of Physics: Conference
  Series}\ }\textbf {\bibinfo {volume} {391}},\ \bibinfo {pages} {012032}
  (\bibinfo {year} {2012})}\BibitemShut {NoStop}%
\bibitem [{\citenamefont {Kummer}\ \emph {et~al.}(2016)\citenamefont {Kummer},
  \citenamefont {Geibel}, \citenamefont {Krellner}, \citenamefont {Zwicknagl},
  \citenamefont {Laubschat}, \citenamefont {Brookes},\ and\ \citenamefont
  {Vyalikh}}]{Kummer-2016}%
  \BibitemOpen
  \bibfield  {author} {\bibinfo {author} {\bibfnamefont {K.}~\bibnamefont
  {Kummer}}, \bibinfo {author} {\bibfnamefont {C.}~\bibnamefont {Geibel}},
  \bibinfo {author} {\bibfnamefont {C.}~\bibnamefont {Krellner}}, \bibinfo
  {author} {\bibfnamefont {G.}~\bibnamefont {Zwicknagl}}, \bibinfo {author}
  {\bibfnamefont {C.}~\bibnamefont {Laubschat}}, \bibinfo {author}
  {\bibfnamefont {N.~B.}\ \bibnamefont {Brookes}}, \ and\ \bibinfo {author}
  {\bibfnamefont {D.~V.}\ \bibnamefont {Vyalikh}},\ }\href@noop {} {\bibfield
  {journal} {\bibinfo  {journal} {arXiv:1612.02169}\ } (\bibinfo {year}
  {2016})}\BibitemShut {NoStop}%
\bibitem [{\citenamefont {Kliemt}\ and\ \citenamefont
  {Cornelius}(2016)}]{Kliemt-2016}%
  \BibitemOpen
  \bibfield  {author} {\bibinfo {author} {\bibfnamefont {K.}~\bibnamefont
  {Kliemt}}\ and\ \bibinfo {author} {\bibfnamefont {K.}~\bibnamefont
  {Cornelius}},\ }\href@noop {} {\bibfield  {journal} {\bibinfo  {journal}
  {arXiv:1612.00753}\ } (\bibinfo {year} {2016})}\BibitemShut {NoStop}%
\bibitem [{\citenamefont {Sarkar}\ \emph {et~al.}(2012)\citenamefont {Sarkar},
  \citenamefont {Khuntia}, \citenamefont {Krellner}, \citenamefont {Geibel},
  \citenamefont {Steglich},\ and\ \citenamefont {Baenitz}}]{Sarkar-2012}%
  \BibitemOpen
  \bibfield  {author} {\bibinfo {author} {\bibfnamefont {R.}~\bibnamefont
  {Sarkar}}, \bibinfo {author} {\bibfnamefont {P.}~\bibnamefont {Khuntia}},
  \bibinfo {author} {\bibfnamefont {C.}~\bibnamefont {Krellner}}, \bibinfo
  {author} {\bibfnamefont {C.}~\bibnamefont {Geibel}}, \bibinfo {author}
  {\bibfnamefont {F.}~\bibnamefont {Steglich}}, \ and\ \bibinfo {author}
  {\bibfnamefont {M.}~\bibnamefont {Baenitz}},\ }\href {\doibase
  10.1103/PhysRevB.85.140409} {\bibfield  {journal} {\bibinfo  {journal} {Phys.
  Rev. B}\ }\textbf {\bibinfo {volume} {85}},\ \bibinfo {pages} {140409}
  (\bibinfo {year} {2012})}\BibitemShut {NoStop}%
\bibitem [{\citenamefont {Baenitz}\ \emph {et~al.}(2013)\citenamefont
  {Baenitz}, \citenamefont {Sarkar}, \citenamefont {Khuntia}, \citenamefont
  {Krellner}, \citenamefont {Geibel},\ and\ \citenamefont
  {Steglich}}]{Baenitz-2013}%
  \BibitemOpen
  \bibfield  {author} {\bibinfo {author} {\bibfnamefont {M.}~\bibnamefont
  {Baenitz}}, \bibinfo {author} {\bibfnamefont {R.}~\bibnamefont {Sarkar}},
  \bibinfo {author} {\bibfnamefont {P.}~\bibnamefont {Khuntia}}, \bibinfo
  {author} {\bibfnamefont {C.}~\bibnamefont {Krellner}}, \bibinfo {author}
  {\bibfnamefont {C.}~\bibnamefont {Geibel}}, \ and\ \bibinfo {author}
  {\bibfnamefont {F.}~\bibnamefont {Steglich}},\ }\href@noop {} {\bibfield
  {journal} {\bibinfo  {journal} {physica status solidi (c)}\ }\textbf
  {\bibinfo {volume} {10}},\ \bibinfo {pages} {540} (\bibinfo {year}
  {2013})}\BibitemShut {NoStop}%
\bibitem [{\citenamefont {Huesges}\ \emph {et~al.}(2015)\citenamefont
  {Huesges}, \citenamefont {Koza}, \citenamefont {Embs}, \citenamefont
  {Fennell}, \citenamefont {Simeoni}, \citenamefont {Geibel}, \citenamefont
  {Krellner},\ and\ \citenamefont {Stockert}}]{Huesges-JPCM-2015}%
  \BibitemOpen
  \bibfield  {author} {\bibinfo {author} {\bibfnamefont {Z.}~\bibnamefont
  {Huesges}}, \bibinfo {author} {\bibfnamefont {M.~M.}\ \bibnamefont {Koza}},
  \bibinfo {author} {\bibfnamefont {J.~P.}\ \bibnamefont {Embs}}, \bibinfo
  {author} {\bibfnamefont {T.}~\bibnamefont {Fennell}}, \bibinfo {author}
  {\bibfnamefont {G.}~\bibnamefont {Simeoni}}, \bibinfo {author} {\bibfnamefont
  {C.}~\bibnamefont {Geibel}}, \bibinfo {author} {\bibfnamefont
  {C.}~\bibnamefont {Krellner}}, \ and\ \bibinfo {author} {\bibfnamefont
  {O.}~\bibnamefont {Stockert}},\ }\href
  {http://stacks.iop.org/1742-6596/592/i=1/a=012083} {\bibfield  {journal}
  {\bibinfo  {journal} {Journal of Physics: Conference Series}\ }\textbf
  {\bibinfo {volume} {592}},\ \bibinfo {pages} {012083} (\bibinfo {year}
  {2015})}\BibitemShut {NoStop}%
\bibitem [{\citenamefont {Kliemt}\ and\ \citenamefont
  {Krellner}(2016)}]{Krellner-2016}%
  \BibitemOpen
  \bibfield  {author} {\bibinfo {author} {\bibfnamefont {K.}~\bibnamefont
  {Kliemt}}\ and\ \bibinfo {author} {\bibfnamefont {C.}~\bibnamefont
  {Krellner}},\ }\href {\doibase
  http://dx.doi.org/10.1016/j.jcrysgro.2016.05.042} {\bibfield  {journal}
  {\bibinfo  {journal} {Journal of Crystal Growth}\ }\textbf {\bibinfo {volume}
  {449}},\ \bibinfo {pages} {129 } (\bibinfo {year} {2016})}\BibitemShut
  {NoStop}%
\bibitem [{\citenamefont {Schenck}(1985)}]{schenck1985muon}%
  \BibitemOpen
  \bibfield  {author} {\bibinfo {author} {\bibfnamefont {A.}~\bibnamefont
  {Schenck}},\ }\href {http://books.google.de/books?id=b6dPAQAAIAAJ} {\emph
  {\bibinfo {title} {Muon spin rotation spectroscopy}}}\ (\bibinfo {year}
  {1985})\BibitemShut {NoStop}%
\bibitem [{\citenamefont {Yaouanc}\ and\ \citenamefont
  {de~R{\'e}otier}(2011)}]{yaouanc2011muon}%
  \BibitemOpen
  \bibfield  {author} {\bibinfo {author} {\bibfnamefont {A.}~\bibnamefont
  {Yaouanc}}\ and\ \bibinfo {author} {\bibfnamefont {P.}~\bibnamefont
  {de~R{\'e}otier}},\ }\href {http://books.google.ch/books?id=nZyicQAACAAJ}
  {\emph {\bibinfo {title} {Muon Spin Rotation, Relaxation, and Resonance:
  Applications to Condensed Matter}}},\ International Series of Monographs on
  Physics\ (\bibinfo  {publisher} {OUP Oxford},\ \bibinfo {year}
  {2011})\BibitemShut {NoStop}%
\bibitem [{\citenamefont {Suter}\ and\ \citenamefont
  {Wojek}(2012)}]{Suter201269}%
  \BibitemOpen
  \bibfield  {author} {\bibinfo {author} {\bibfnamefont {A.}~\bibnamefont
  {Suter}}\ and\ \bibinfo {author} {\bibfnamefont {B.}~\bibnamefont {Wojek}},\
  }\href {\doibase http://dx.doi.org/10.1016/j.phpro.2012.04.042} {\bibfield
  {journal} {\bibinfo  {journal} {Physics Procedia}\ }\textbf {\bibinfo
  {volume} {30}},\ \bibinfo {pages} {69 } (\bibinfo {year} {2012})}\BibitemShut
  {NoStop}%
\bibitem [{\citenamefont {Barsov}\ \emph {et~al.}(1991)\citenamefont {Barsov},
  \citenamefont {Getalov}, \citenamefont {Ginsburg}, \citenamefont {Koptev},
  \citenamefont {Kruglov}, \citenamefont {Kuzmin}, \citenamefont {Maleyev},
  \citenamefont {Maltsev}, \citenamefont {Mikirtychyants}, \citenamefont
  {Tarasov}, \citenamefont {Shcherbakov}, \citenamefont {Grebinnik},
  \citenamefont {Duginov}, \citenamefont {Lazarev}, \citenamefont {Olshevski},
  \citenamefont {Shilov}, \citenamefont {Zhukov}, \citenamefont {Gurevich},
  \citenamefont {Kirillov}, \citenamefont {Klimov}, \citenamefont {Nikolski},
  \citenamefont {Pirogov}, \citenamefont {Ponomarev},\ and\ \citenamefont
  {Suetin}}]{Barsov-1991}%
  \BibitemOpen
  \bibfield  {author} {\bibinfo {author} {\bibfnamefont {S.}~\bibnamefont
  {Barsov}}, \bibinfo {author} {\bibfnamefont {A.}~\bibnamefont {Getalov}},
  \bibinfo {author} {\bibfnamefont {S.}~\bibnamefont {Ginsburg}}, \bibinfo
  {author} {\bibfnamefont {V.}~\bibnamefont {Koptev}}, \bibinfo {author}
  {\bibfnamefont {S.}~\bibnamefont {Kruglov}}, \bibinfo {author} {\bibfnamefont
  {L.}~\bibnamefont {Kuzmin}}, \bibinfo {author} {\bibfnamefont
  {S.}~\bibnamefont {Maleyev}}, \bibinfo {author} {\bibfnamefont
  {E.}~\bibnamefont {Maltsev}}, \bibinfo {author} {\bibfnamefont
  {S.}~\bibnamefont {Mikirtychyants}}, \bibinfo {author} {\bibfnamefont
  {N.}~\bibnamefont {Tarasov}}, \bibinfo {author} {\bibfnamefont
  {G.}~\bibnamefont {Shcherbakov}}, \bibinfo {author} {\bibfnamefont
  {V.}~\bibnamefont {Grebinnik}}, \bibinfo {author} {\bibfnamefont
  {V.}~\bibnamefont {Duginov}}, \bibinfo {author} {\bibfnamefont
  {A.}~\bibnamefont {Lazarev}}, \bibinfo {author} {\bibfnamefont
  {V.}~\bibnamefont {Olshevski}}, \bibinfo {author} {\bibfnamefont
  {S.}~\bibnamefont {Shilov}}, \bibinfo {author} {\bibfnamefont
  {V.}~\bibnamefont {Zhukov}}, \bibinfo {author} {\bibfnamefont
  {I.}~\bibnamefont {Gurevich}}, \bibinfo {author} {\bibfnamefont
  {B.}~\bibnamefont {Kirillov}}, \bibinfo {author} {\bibfnamefont
  {A.}~\bibnamefont {Klimov}}, \bibinfo {author} {\bibfnamefont
  {B.}~\bibnamefont {Nikolski}}, \bibinfo {author} {\bibfnamefont
  {A.}~\bibnamefont {Pirogov}}, \bibinfo {author} {\bibfnamefont
  {A.}~\bibnamefont {Ponomarev}}, \ and\ \bibinfo {author} {\bibfnamefont
  {V.}~\bibnamefont {Suetin}},\ }\href {\doibase 10.1007/BF02396171} {\bibfield
   {journal} {\bibinfo  {journal} {Hyperfine Interactions}\ }\textbf {\bibinfo
  {volume} {64}},\ \bibinfo {pages} {415} (\bibinfo {year} {1991})}\BibitemShut
  {NoStop}%
\bibitem [{\citenamefont {Kornilov}\ and\ \citenamefont
  {Pomjakushin}(1991)}]{Kornilov-1991}%
  \BibitemOpen
  \bibfield  {author} {\bibinfo {author} {\bibfnamefont {E.}~\bibnamefont
  {Kornilov}}\ and\ \bibinfo {author} {\bibfnamefont {V.}~\bibnamefont
  {Pomjakushin}},\ }\href {\doibase 10.1016/0375-9601(91)90959-C} {\bibfield
  {journal} {\bibinfo  {journal} {Physics Letters A}\ }\textbf {\bibinfo
  {volume} {153}},\ \bibinfo {pages} {364 } (\bibinfo {year}
  {1991})}\BibitemShut {NoStop}%
\bibitem [{\citenamefont {MacLaughlin}\ \emph {et~al.}(2004)\citenamefont
  {MacLaughlin}, \citenamefont {Heffner}, \citenamefont {Bernal}, \citenamefont
  {Ishida}, \citenamefont {Sonier}, \citenamefont {Nieuwenhuys}, \citenamefont
  {Maple},\ and\ \citenamefont {Stewart}}]{MacLaughlin-time-field-scaling}%
  \BibitemOpen
  \bibfield  {author} {\bibinfo {author} {\bibfnamefont {D.~E.}\ \bibnamefont
  {MacLaughlin}}, \bibinfo {author} {\bibfnamefont {R.~H.}\ \bibnamefont
  {Heffner}}, \bibinfo {author} {\bibfnamefont {O.~O.}\ \bibnamefont {Bernal}},
  \bibinfo {author} {\bibfnamefont {K.}~\bibnamefont {Ishida}}, \bibinfo
  {author} {\bibfnamefont {J.~E.}\ \bibnamefont {Sonier}}, \bibinfo {author}
  {\bibfnamefont {G.~J.}\ \bibnamefont {Nieuwenhuys}}, \bibinfo {author}
  {\bibfnamefont {M.~B.}\ \bibnamefont {Maple}}, \ and\ \bibinfo {author}
  {\bibfnamefont {G.~R.}\ \bibnamefont {Stewart}},\ }\href
  {http://stacks.iop.org/0953-8984/16/i=40/a=005} {\bibfield  {journal}
  {\bibinfo  {journal} {Journal of Physics: Condensed Matter}\ }\textbf
  {\bibinfo {volume} {16}},\ \bibinfo {pages} {S4479} (\bibinfo {year}
  {2004})}\BibitemShut {NoStop}%
\bibitem [{\citenamefont {Michor}\ \emph {et~al.}(2006)\citenamefont {Michor},
  \citenamefont {Adroja}, \citenamefont {Bauer}, \citenamefont {Bewley},
  \citenamefont {Dobozanov}, \citenamefont {Hillier}, \citenamefont {Hilscher},
  \citenamefont {Killer}, \citenamefont {Koza}, \citenamefont {Manalo},
  \citenamefont {Manuel}, \citenamefont {Reissner}, \citenamefont {Rogl},
  \citenamefont {Rotter},\ and\ \citenamefont {Scheidt}}]{Michor2006640}%
  \BibitemOpen
  \bibfield  {author} {\bibinfo {author} {\bibfnamefont {H.}~\bibnamefont
  {Michor}}, \bibinfo {author} {\bibfnamefont {D.}~\bibnamefont {Adroja}},
  \bibinfo {author} {\bibfnamefont {E.}~\bibnamefont {Bauer}}, \bibinfo
  {author} {\bibfnamefont {R.}~\bibnamefont {Bewley}}, \bibinfo {author}
  {\bibfnamefont {D.}~\bibnamefont {Dobozanov}}, \bibinfo {author}
  {\bibfnamefont {A.}~\bibnamefont {Hillier}}, \bibinfo {author} {\bibfnamefont
  {G.}~\bibnamefont {Hilscher}}, \bibinfo {author} {\bibfnamefont
  {U.}~\bibnamefont {Killer}}, \bibinfo {author} {\bibfnamefont
  {M.}~\bibnamefont {Koza}}, \bibinfo {author} {\bibfnamefont {S.}~\bibnamefont
  {Manalo}}, \bibinfo {author} {\bibfnamefont {P.}~\bibnamefont {Manuel}},
  \bibinfo {author} {\bibfnamefont {M.}~\bibnamefont {Reissner}}, \bibinfo
  {author} {\bibfnamefont {P.}~\bibnamefont {Rogl}}, \bibinfo {author}
  {\bibfnamefont {M.}~\bibnamefont {Rotter}}, \ and\ \bibinfo {author}
  {\bibfnamefont {E.-W.}\ \bibnamefont {Scheidt}},\ }\href {\doibase
  http://dx.doi.org/10.1016/j.physb.2006.01.345} {\bibfield  {journal}
  {\bibinfo  {journal} {Physica B: Condensed Matter}\ }\textbf {\bibinfo
  {volume} {378–380}},\ \bibinfo {pages} {640 } (\bibinfo {year}
  {2006})}\BibitemShut {NoStop}%
\bibitem [{\citenamefont {Hayano}\ \emph {et~al.}(1979)\citenamefont {Hayano},
  \citenamefont {Uemura}, \citenamefont {Imazato}, \citenamefont {Nishida},
  \citenamefont {Yamazaki},\ and\ \citenamefont
  {Kubo}}]{PhysRevB.20.850-Hayana}%
  \BibitemOpen
  \bibfield  {author} {\bibinfo {author} {\bibfnamefont {R.~S.}\ \bibnamefont
  {Hayano}}, \bibinfo {author} {\bibfnamefont {Y.~J.}\ \bibnamefont {Uemura}},
  \bibinfo {author} {\bibfnamefont {J.}~\bibnamefont {Imazato}}, \bibinfo
  {author} {\bibfnamefont {N.}~\bibnamefont {Nishida}}, \bibinfo {author}
  {\bibfnamefont {T.}~\bibnamefont {Yamazaki}}, \ and\ \bibinfo {author}
  {\bibfnamefont {R.}~\bibnamefont {Kubo}},\ }\href {\doibase
  10.1103/PhysRevB.20.850} {\bibfield  {journal} {\bibinfo  {journal} {Phys.
  Rev. B}\ }\textbf {\bibinfo {volume} {20}},\ \bibinfo {pages} {850} (\bibinfo
  {year} {1979})}\BibitemShut {NoStop}%
\bibitem [{\citenamefont {Moriya}(1985)}]{Moriya-1985}%
  \BibitemOpen
  \bibfield  {author} {\bibinfo {author} {\bibfnamefont {T.}~\bibnamefont
  {Moriya}},\ }\href@noop {} {\  (\bibinfo {year} {1985})}\BibitemShut
  {NoStop}%
\bibitem [{\citenamefont {Hatatani}\ and\ \citenamefont
  {Moriya}(1995)}]{Hatatani-1995}%
  \BibitemOpen
  \bibfield  {author} {\bibinfo {author} {\bibfnamefont {M.}~\bibnamefont
  {Hatatani}}\ and\ \bibinfo {author} {\bibfnamefont {T.}~\bibnamefont
  {Moriya}},\ }\href@noop {} {\bibfield  {journal} {\bibinfo  {journal}
  {Journal of the Physical Society of Japan}\ }\textbf {\bibinfo {volume}
  {64}},\ \bibinfo {pages} {3434} (\bibinfo {year} {1995})}\BibitemShut
  {NoStop}%
\bibitem [{\citenamefont {Stewart}(2001)}]{RevModPhys.73.797}%
  \BibitemOpen
  \bibfield  {author} {\bibinfo {author} {\bibfnamefont {G.~R.}\ \bibnamefont
  {Stewart}},\ }\href {\doibase 10.1103/RevModPhys.73.797} {\bibfield
  {journal} {\bibinfo  {journal} {Rev. Mod. Phys.}\ }\textbf {\bibinfo {volume}
  {73}},\ \bibinfo {pages} {797} (\bibinfo {year} {2001})}\BibitemShut
  {NoStop}%
\bibitem [{\citenamefont {Carretta}\ \emph {et~al.}(2009)\citenamefont
  {Carretta}, \citenamefont {Pasero}, \citenamefont {Giovannini},\ and\
  \citenamefont {Baines}}]{Carretta-2009}%
  \BibitemOpen
  \bibfield  {author} {\bibinfo {author} {\bibfnamefont {P.}~\bibnamefont
  {Carretta}}, \bibinfo {author} {\bibfnamefont {R.}~\bibnamefont {Pasero}},
  \bibinfo {author} {\bibfnamefont {M.}~\bibnamefont {Giovannini}}, \ and\
  \bibinfo {author} {\bibfnamefont {C.}~\bibnamefont {Baines}},\ }\href
  {\doibase 10.1103/PhysRevB.79.020401} {\bibfield  {journal} {\bibinfo
  {journal} {Phys. Rev. B}\ }\textbf {\bibinfo {volume} {79}},\ \bibinfo
  {pages} {020401} (\bibinfo {year} {2009})}\BibitemShut {NoStop}%
\bibitem [{\citenamefont {Toll}(1956)}]{Toll-dissipation-theorem}%
  \BibitemOpen
  \bibfield  {author} {\bibinfo {author} {\bibfnamefont {J.~S.}\ \bibnamefont
  {Toll}},\ }\href {\doibase 10.1103/PhysRev.104.1760} {\bibfield  {journal}
  {\bibinfo  {journal} {Phys. Rev.}\ }\textbf {\bibinfo {volume} {104}},\
  \bibinfo {pages} {1760} (\bibinfo {year} {1956})}\BibitemShut {NoStop}%
\bibitem [{\citenamefont {Ishida}\ \emph {et~al.}(2003)\citenamefont {Ishida},
  \citenamefont {MacLaughlin}, \citenamefont {Young}, \citenamefont {Okamoto},
  \citenamefont {Kawasaki}, \citenamefont {Kitaoka}, \citenamefont
  {Nieuwenhuys}, \citenamefont {Heffner}, \citenamefont {Bernal}, \citenamefont
  {Higemoto}, \citenamefont {Koda}, \citenamefont {Kadono}, \citenamefont
  {Trovarelli}, \citenamefont {Geibel},\ and\ \citenamefont
  {Steglich}}]{Ishida-musr-YbRh2Si2}%
  \BibitemOpen
  \bibfield  {author} {\bibinfo {author} {\bibfnamefont {K.}~\bibnamefont
  {Ishida}}, \bibinfo {author} {\bibfnamefont {D.~E.}\ \bibnamefont
  {MacLaughlin}}, \bibinfo {author} {\bibfnamefont {B.-L.}\ \bibnamefont
  {Young}}, \bibinfo {author} {\bibfnamefont {K.}~\bibnamefont {Okamoto}},
  \bibinfo {author} {\bibfnamefont {Y.}~\bibnamefont {Kawasaki}}, \bibinfo
  {author} {\bibfnamefont {Y.}~\bibnamefont {Kitaoka}}, \bibinfo {author}
  {\bibfnamefont {G.~J.}\ \bibnamefont {Nieuwenhuys}}, \bibinfo {author}
  {\bibfnamefont {R.~H.}\ \bibnamefont {Heffner}}, \bibinfo {author}
  {\bibfnamefont {O.~O.}\ \bibnamefont {Bernal}}, \bibinfo {author}
  {\bibfnamefont {W.}~\bibnamefont {Higemoto}}, \bibinfo {author}
  {\bibfnamefont {A.}~\bibnamefont {Koda}}, \bibinfo {author} {\bibfnamefont
  {R.}~\bibnamefont {Kadono}}, \bibinfo {author} {\bibfnamefont
  {O.}~\bibnamefont {Trovarelli}}, \bibinfo {author} {\bibfnamefont
  {C.}~\bibnamefont {Geibel}}, \ and\ \bibinfo {author} {\bibfnamefont
  {F.}~\bibnamefont {Steglich}},\ }\href {\doibase 10.1103/PhysRevB.68.184401}
  {\bibfield  {journal} {\bibinfo  {journal} {Phys. Rev. B}\ }\textbf {\bibinfo
  {volume} {68}},\ \bibinfo {pages} {184401} (\bibinfo {year}
  {2003})}\BibitemShut {NoStop}%
\end{thebibliography}%
\end{document}